\documentclass[10pt]{article}

\usepackage[margin=0.70in]{geometry}
\usepackage{graphicx}
\usepackage{xcolor}
\usepackage{booktabs}
\usepackage{amsmath}
\usepackage{amssymb}
\usepackage{tabularx}
\usepackage{url}
\definecolor{linkblue}{RGB}{0,102,204}
\usepackage[
    colorlinks=true,
    linkcolor=black,
    citecolor=black,
    urlcolor=linkblue
]{hyperref}

\usepackage{enumitem}
\usepackage{microtype}
\usepackage{textcomp}

\newcommand{\repo}{\url{https://github.com/abhishekhanchate/nsf-fmrg-data-challenge}}
\newcommand{\datasetdoi}{\href{https://doi.org/10.5281/zenodo.21285367}{doi:10.5281/zenodo.21285367}}

\graphicspath{{figures/}}

\title{NSF Future Manufacturing Data Challenge: A Multimodal DED Dataset for Probabilistic Representation and Prediction of Laser-Track Geometry}

\author{Abhishek Hanchate$^{\ast}$ \quad Himanshu Balhara \quad Satish T. S. Bukkapatnam\\
\small Wm Michael Barnes '64 Department of Industrial and Systems Engineering, Texas A\&M University\\
\small $^{\ast}$Corresponding author: \href{mailto:abhishek.hanchate@tamu.edu}{abhishek.hanchate@tamu.edu}}

\begin{document}
\maketitle

\begin{abstract}
We introduce a multimodal directed energy deposition (DED) dataset for predicting the probabilistic local geometric variation of single laser tracks produced on stainless-steel 316L substrates. The dataset supports the NSF Future Manufacturing Data Challenge and contains three complementary modalities: in-situ thermal image sequences from a Stratonics ThermaViz melt-pool sensor, scanning electron microscopy (SEM) images acquired using a Zeiss EVO MA10 system, and full-field height maps acquired using a Bruker ContourGT-K white-light 3D optical profilometer. Each experiment is a bead-on-plate scan at one of four laser powers, 200, 300, 350, and 400 W, with a fixed scan speed of 10 mm/s. The release includes a multimodal coordinate convention linking thermal, SEM, and height-map measurements over a common physical 20--100 mm window, with the raw dataset available on \href{https://doi.org/10.5281/zenodo.21285367}{Zenodo} and participant-facing notebooks, reusable code, and documentation available on \href{https://github.com/abhishekhanchate/nsf-fmrg-data-challenge}{GitHub}.
\end{abstract}

\section{Motivation}
Laser-based DED produces spatially varying tracks whose final geometry depends on both process-driven thermal history and the local state of the substrate. This framing follows related studies on multimodal sensing and spatial alignment in DED \cite{hira2024,porosity2025}, morphology representation from height maps \cite{hanchate2026morphology}, and ripple-driven surface variation in DED 316L components \cite{balhara2023ripples}. In practice, track formation is not perfectly uniform along the scan direction: local melt-pool dynamics, heat dissipation, and pre-existing surface morphology can all influence the final geometry. The present challenge is designed to make those local dependencies explicit.

The dataset is designed to support models that combine process-driven and substrate-driven sources of variation. Thermal frames provide a time-resolved view of the moving melt pool and its thermal neighborhood. SEM images provide surface context around the track and are intended to describe local valleys, ridges, ripples, and other substrate features. Bruker profilometry provides post-process full-field height maps from which local geometric descriptors can be extracted. By combining these modalities, the challenge encourages participants to move beyond a single global width value and instead model local geometry as a spatially varying quantity.

\section{Challenge Task}
Given thermal frames $T_{i,t}\in\mathbb{R}^{400\times 400}$, an SEM context image $S_i$, and a final height map $Y_i$, participants are asked to model a local geometry descriptor
\begin{equation}
    p\left(g_i(x) \mid T_{i,t-k:t+k}, S_i, x\right),
\end{equation}
where $x$ is the physical scan-direction coordinate and $g_i(x)$ may represent local width, left/right boundary position, contour deviation, edge roughness, waviness, or a vector of local geometric descriptors. The primary expected target is local width variation extracted from the height map, but the dataset also supports alternative definitions of local boundary and contour irregularity.

SEM imagery should be used only to characterize surrounding substrate morphology. The processed track region should be masked or otherwise excluded when SEM features are used, so that models do not leak information from the final track appearance into the prediction target. Participants may use hand-engineered descriptors, learned features, sequence models, probabilistic regressors, or multimodal fusion models, but should report how the input information is aligned to physical track coordinates.

\begin{figure}[t]
\centering
\includegraphics[width=0.98\linewidth]{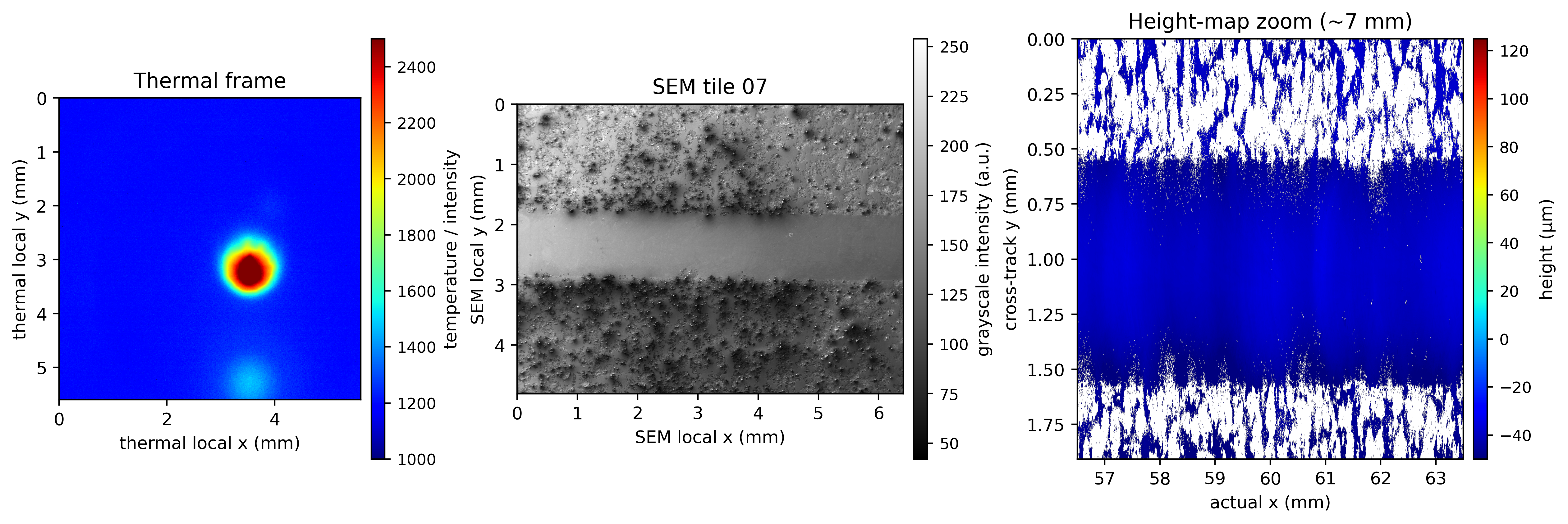}
\caption{Representative modalities in the challenge. Left: a thermal frame from the extracted 20--100 mm scan window visualized using a fixed jet scale from 1000 to 2500. Each thermal frame is $400\times400$ pixels with 14 $\mu$m/pixel resolution, corresponding to a 5.6 mm $\times$ 5.6 mm field of view; the melt pool occupies a smaller central region but drives process-dependent variation through its size, shape, intensity, asymmetry, and cooling tail. Middle: SEM tile 07, used as a local substrate-morphology example without stitching. Each SEM tile spans approximately 6.41 mm in the scan direction. Right: a zoomed Bruker/Wyko height-map region over an approximately 7 mm physical window, shown in corrected physical coordinates with axes in millimeters and height in micrometers.}
\label{fig:modalities}
\end{figure}

Representative examples of the three released modalities are shown in Fig.~\ref{fig:modalities}. In particular, the thermal frame illustrates the local melt-pool neighborhood, the SEM tile illustrates the surrounding substrate morphology, and the height-map crop illustrates the type of local geometric variation that may be used as a prediction target.

\section{Experimental Setting}
\begin{figure}[t]
\centering
\includegraphics[width=0.98\linewidth]{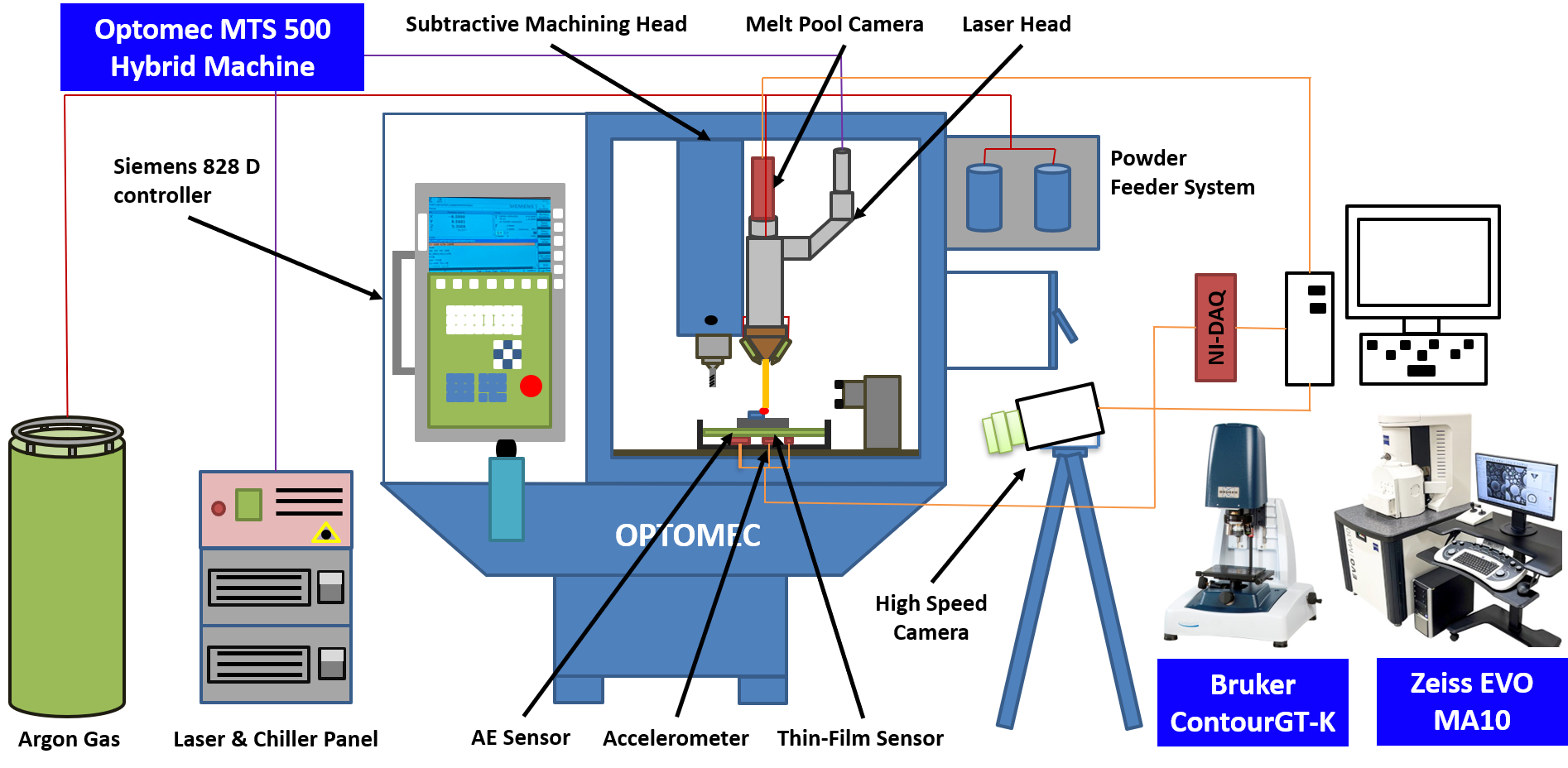}
\caption{Experimental setup used for data generation and characterization. The dataset was collected on an Optomec MTS 500 hybrid machine. In-situ thermal data were obtained during the bead-on-plate scans, while post-process geometry and morphology information were acquired using a Bruker ContourGT-K white-light 3D optical profilometer and a Zeiss EVO MA10 SEM system.}
\label{fig:setup}
\end{figure}

The experimental setup is summarized in Fig.~\ref{fig:setup}. All experiments were performed using an Optomec LENS\textsuperscript{\textregistered} MTS 500 directed energy deposition smart hybrid platform at Texas A\&M University. Single-track bead-on-plate scans were conducted on SS316L substrates without powder feed to isolate laser--material interaction and resolidification behavior from powder-flow effects. Four laser power settings were investigated, 200, 300, 350, and 400 W, at a fixed scan speed of 10 mm/s. The released files use track IDs 8, 10, 14, and 21.

The common analysis window is approximately 80 mm, corresponding to actual part coordinates 20--100 mm. Although the programmed scan is longer, the released synchronized window focuses on the region for which thermal, SEM, and height-map information can be placed into a common physical coordinate convention. The dataset therefore emphasizes local geometry prediction over an aligned physical interval rather than whole-file processing without coordinate correction.

\section{Modality Details and Physical Coordinates}
Table~\ref{tab:modalities} summarizes the key units, physical scale factors, and coordinate conventions used in the released notebooks.

\begin{table}[t]
\centering
\caption{Summary of released modalities and physical conventions.}
\small
\begin{tabularx}{\linewidth}{l l l X}
\toprule
Modality & File type & Native scale & Physical interpretation \\
\midrule
Thermal & \texttt{.mat} & $400\times400$ pixels, 50 fps & 14 $\mu$m/pixel, 5.6 mm $\times$ 5.6 mm field of view. The 20--100 mm segment is extracted backward from the detected laser-stop frame. \\
SEM & \texttt{.tif} tiles & 13--14 tiles per track & Tile 01 corresponds to the 100 mm side, while the highest-numbered tile corresponds to the 20 mm side. Tile 07 is used as a representative local surface-context example. \\
Height map & Wyko ASCII & $480\times X$ grid & $x,y$ are stored in mm and $z$ in nm. The code converts $z$ to mm or $\mu$m and remaps raw ASC direction into increasing 20--100 mm actual coordinate order. \\
\bottomrule
\end{tabularx}
\label{tab:modalities}
\end{table}

\paragraph{Thermal images.}
The thermal camera records the melt-pool region at 50 frames per second. At a scan speed of 10 mm/s, consecutive active frames correspond to
\begin{equation}
\Delta x = \frac{10~\mathrm{mm/s}}{50~\mathrm{frames/s}} = 0.2~\mathrm{mm/frame}.
\end{equation}
Each frame is $400\times400$ pixels with approximately 14 $\mu$m/pixel resolution, giving a 5.6 mm $\times$ 5.6 mm field of view. The melt pool is a smaller region near the center of this field of view. Its local size, shape, intensity distribution, asymmetry, and trailing thermal field are expected to encode process-driven variation that influences the final track geometry.

\paragraph{SEM images.}
The SEM images provide local surface-morphology context. For each track, the \texttt{PlainImages} folder contains the unscaled tile images. Track IDs 8, 10, and 14 have 13 tiles, while Track 21 has 14 tiles. Tile 01 corresponds to the 100 mm side of the physical coordinate convention, while the highest-numbered tile corresponds to the 20 mm side. In the starter material, SEM images are shown as individual tiles rather than stitched mosaics to avoid introducing registration artifacts into the participant-facing baseline. As summarized in Table~\ref{tab:modalities}, each tile spans approximately 6.41 mm in the scan direction, and a representative example is shown in Fig.~\ref{fig:modalities}.

\paragraph{Height maps.}
Post-process height maps are acquired using a Bruker ContourGT-K white-light 3D optical profilometer and stored in Wyko ASCII format. The ASCII rows contain $x$, $y$, and $z$ values, where $x$ and $y$ are in millimeters and $z$ is in nanometers. The code converts $z$ to millimeters internally and typically displays height in micrometers. The raw ASC direction is opposite the desired part-coordinate direction: local ASC $x=0$ corresponds to the physical 100 mm side. The starter loader therefore sorts columns into increasing actual coordinate order and crops the common 20--100 mm window. Key file-level metadata for the released height maps and thermal windows are listed in Table~\ref{tab:data}.

\begin{table}[t]
\centering
\caption{Verified metadata used in the starter notebooks. Thermal intervals are Python-style half-open frame ranges.}
\small
\begin{tabular}{lcccccc}
\toprule
Track & Raw & Laser-on & Extracted & Header $X$ & Loaded $Z[y,x]$ & NaN frac. \\
\midrule
8  & 929  & 261--737 & 337--737 & 20025 & $480\times 20025$ & 0.369 \\
10 & 961  & 265--741 & 341--741 & 20224 & $480\times 20091$ & 0.516 \\
14 & 976  & 309--785 & 385--785 & 19724 & $480\times 19724$ & 0.511 \\
21 & 1012 & 253--729 & 329--729 & 19909 & $480\times 19909$ & 0.555 \\
\bottomrule
\end{tabular}
\label{tab:data}
\end{table}

\section{Suggested Processing and Evaluation}
The repository includes a participant-facing notebook for loading thermal, SEM, and Bruker/Wyko data; an organizer/testing Colab notebook for exporting the final thermal segment, thermal videos, and figures; and reusable loader functions. Additional diagnostic plots include fixed-range height-map visualizations and display-only manual tilt corrections for selected tracks. These tools are intended as examples rather than a prescribed modeling pipeline.

No hidden leaderboard labels are used in the current release. Because only four track conditions are included, development should use cross-track validation, with Track 21 preferably treated as a held-out test case. This choice is conservative because Track 21 has a less complete post-process profilometry capture due to profilometer acquisition failure, making it useful for testing robustness to incomplete height-map coverage rather than tuning preprocessing choices. Candidate metrics include mean absolute error for local width, boundary-position error, negative log likelihood or continuous ranked probability score for probabilistic predictions, and calibration error for uncertainty-aware outputs.

\section{Release and Citation}
The complete raw multimodal dataset is publicly available through Zenodo at \datasetdoi. Participant-facing loading and visualization notebooks, reusable processing functions, and supporting documentation are available in the accompanying GitHub repository at \repo. Challenge participants and subsequent users of the dataset should cite this dataset paper and the archived Zenodo release.

\paragraph{Acknowledgment.}
This competition and associated material are based upon work supported by the National Science Foundation under Grant Number FMRG-2328395.

\end{document}